\documentclass[fleqn,usenatbib]{mnras}

\usepackage{newtxtext,newtxmath}
\usepackage[T1]{fontenc}

\DeclareRobustCommand{\VAN}[3]{#2}
\let\VANthebibliography\thebibliography
\def\thebibliography{\DeclareRobustCommand{\VAN}[3]{##3}\VANthebibliography}

\usepackage{slashed}
\usepackage{amsmath}
\usepackage{physics}
\usepackage{color}
\usepackage[dvipsnames]{xcolor}
\usepackage{graphicx}
\usepackage{wrapfig}
\usepackage{threeparttable}

\newcommand{\dq}[1]{\ifmmode d_{#1}^{(2)} \else $d_{#1}^{(2)}$ \fi}
\newcommand{\dme}{\dq{m_e}}
\newcommand{\Ncurl}{\mathcal{N}}

\makeatletter
\newcommand{\tnotelink}{%
  \refstepcounter{footnote}% create a numbered anchor
  \hyperlink{footnote@\thefootnote}{\thefootnote}% clickable mark
}
\newcommand{\tablenotelabel}{%
  \hypertarget{footnote@\thefootnote}{}% place link target
  \thefootnote
}
\makeatother

\title[Ultralight scalars and the WD MRR]{New ultralight scalar particles and the mass-radius relation of white dwarfs -- the important role of Sirius B}

\author[K. Bartnick et al.]{
Kai Bartnick,$^{1}$\thanks{E-mail: kai.bartnick@tum.de}
Detlev Koester,$^{2}$
Rolf-Peter Kudritzki,$^{3,4}$
Konstantin Springmann,$^{5}$
Stefan Stelzl$^{6}$
\newauthor
and Andreas Weiler$^{1}$
\\
$^{1}$Technische Universität München, Physik-Department, James-Franck-Strasse 1, 85748 Garching, Germany\\
$^{2}$Institut für Theoretische Physik und Astrophysik, University of Kiel, 24098 Kiel, Germany\\
$^{3}$Institute for Astronomy, University of Hawaii, 2680 Woodlawn Drive, Honolulu, HI 96822, USA\\
$^{4}$Universitäts-Sternwarte, Fakultät für Physik, Ludwig-Maximilians Universität München, Scheinerstr. 1, D-81679 München, Germany\\
$^{5}$Department of Particle Physics and Astrophysics, Weizmann Institute of Science, Rehovot, Israel 7610001\\
$^{6}$Institute of Physics, Theoretical Particle Physics Laboratory, \'Ecole Polytechnique F\'ed\'erale de Lausanne, CH-1015 Lausanne, Switzerland
}

\date{Accepted 2025 October 12. Received 2025 October 6; in original form 2025 September 3}

\pubyear{2025}

\begin{document}
\label{firstpage}
\pagerange{\pageref{firstpage}--\pageref{lastpage}}
\maketitle

\begin{abstract}
We present the equation of state for two classes of new ultralight particles, a scalar field coupling to electrons and a light $\mathbb{Z}_\mathcal{N}$ QCD axion field coupling to nucleons. Both are potential candidates for dark matter. Using the scalar modified equations of state, we calculate models for white dwarf stars and compare their radii and masses with observed mass-radius data. The comparison results in stringent constraints on the masses of the particles and the coupling parameters. For a wide range of particle masses and coupling parameters, constraints from the white dwarf equation of state surpass existing limits, outperforming also dedicated laboratory searches.
The remarkable accuracy of modern white-dwarf mass–radius relation data, exemplified by Sirius B, now allows stringent tests of dense-matter physics and constraints on new particle scenarios.
\end{abstract}


%%%%%%%%%%%%%%%%%%%%%%%%%%%%%%%%%%%%%%%%%%%%%%%%%%

%%%%%%%%%%%%%%%%% BODY OF PAPER %%%%%%%%%%%%%%%%%%

\section{Introduction}

The discovery of a physical companion to Sirius by Friedrich Wilhelm Bessel in 1841, and the similarity of the spectra of both components in spite of a difference in brightness by many orders of magnitude \citep{Adams1915}, opened a new chapter of stellar astrophysics. The unusual combination of mass and radius of the faint companion Sirius B of the brightest star in the sky led to the concept of degenerate matter and a new equation of state that explained the properties of white dwarfs as end products of stellar evolution \citep{Anderson1929,Stoner01051930,Chandrasekhar31}. The resulting mass-radius relation of white dwarfs (MRR) has become a key ingredient in each textbook of stellar physics. As shown in this paper, the MRR  also provides a unique means of constraining the physics of new particles beyond the Standard Model of particle physics, and Sirius B will play a crucial role in this constraint.

The Standard Model of particle physics (SM) gives an excellent description of many phenomena observed in nature.
Still, several problems and shortcomings motivate us to look for physics that goes beyond.

Most prominently, there is the problem of dark matter, for which the SM has no candidate. Historically, dark matter searches focused on heavy, GeV-scale candidates. However, after decades of null results from indirect dark matter searches and direct detection experiments, particle theorists started to broaden the mass range they considered for their models. A particularly active field looks at ultralight dark matter (ULDM), also referred to as weakly interacting sub-eV particles (WISPs) \citep[see e.g.][]{Preskill:1982cy,Abbott:1982af,Dine:1982ah,Arvanitaki:2014faa}. Instead of hiding dark matter at high energies or masses, detection is avoided by adding very light bosonic particles characterized by minuscule interactions with the Standard Model.
In the end, this can be described by adding a new species of light scalar particles to the SM.

Another theoretical challenge for the Standard Model is the strong CP problem, which describes the violation of charge conjugation (C) and parity (P) symmetry: 
while the electro-weak interaction has an order one CP phase, no violation of CP symmetry has been detected in processes involving only the strong interaction.
There is no explanation for this puzzle within the Standard Model, however, a very attractive solution is a new dynamical field, the bosonic axion \citep{Peccei:1977hh, Peccei:1977ur, Wilczek:1977pj, Weinberg:1977ma}, which is also a viable dark matter candidate. Particles with similar properties, albeit not solving the strong CP problem, are also common in string theory and many other higher-dimensional models \citep{Arvanitaki:2009fg}. These particles are commonly referred to as axion-like particles (ALPs).

Cosmology and the extragalactic distance scale provide an additional motivation to consider new particles beyond the Standard Model. Measurements of the Hubble parameter $H_0$ from the cosmic microwave background compared to the local Universe standard candle approach are incompatible at a high level of statistical significance \citep{Riess2022}. 
Since this tension seems to only increase with new observations \citep{Breuval2024, Riess2024, Kudritzki2024, Adame2025}, increasingly new physics models are used to explain the discrepancy. Some of those also rely on the addition of new scalar fields or require a changing electron mass (see \citealt{DiValentino2021} and \citealt{Schoeneberg2022} for an overview).
\\
 
Typically, these light scalar particles have extremely weak interactions with the Standard Model.
The lighter they are, the weaker are their interactions, and these particles are therefore able to avoid most collider constraints.
Nevertheless, laboratory experiments looking for fifth forces, violations of the equivalence principle, or oscillating fundamental constants are searching for these particles.
For light scalar particles, these experiments produce strict bounds for linearly coupled interactions (see e.g.~\citep{Schlamminger:2007ht,Lee:2020zjt,Tan:2020vpf}) but are significantly less constraining for quadratically coupled interactions.
Additionally, bounds from oscillating fundamental constants only apply if the scalar is responsible for a significant portion of the dark matter energy density in the Universe \citep{Arvanitaki:2014faa,Brzeminski:2020uhm,Stadnik:2014tta,DeRocco:2018jwe}. 

Astrophysics offers alternative approaches to constrain the properties of these scalar particles.
The best-known example is based on stellar cooling: 
if additional light particles couple weakly to Standard Model matter, they get produced in stars and stellar remnants.
Due to the weak coupling, they can easily escape and thus provide an additional energy loss channel for the star.
Thus, stellar cooling in connection with stellar evolution data can be used to constrain these couplings \citep{Raffelt:1996wa,Springmann:2024mjp}.
Considering, for example, QCD axions, generically, there are axion-nucleon couplings, which lead to axion production via nucleon bremsstrahlung. 
Requiring neutron stars not to cool faster than observed yields leading constraints on QCD axions \citep{Buschmann:2021juv,Springmann:2024ret}.

In this work, we will focus on a different approach to constraining particle physics from stars.
Recently \cite{Hook:2017psm} have shown that new scalar fields with quadratic coupling to the Standard Model can develop a classical expectation value within stars - similarly to the Higgs mechanism, but localized within the star (see Fig.~\ref{fig:sourcingSketch}), a phenomenon that might occur for the QCD axion \citep{Balkin:2020dsr}.
In scalar tensor theories of modified gravity, a similar phenomenon is known as scalarization \citep{Damour:1993hw,Doneva:2022ewd,Ramazanoglu:2016kul,Staykov:2018hhc}.
This scalar field profile, especially the part that leaks out of the star can have drastic effects, see \cite{Hook:2017psm,Zhang:2021mks} and \cite{Balkin:2021wea,Balkin:2021zfd}. 
Moreover, the expectation value backreacts on the stellar structure and can lead to a new ground state (NGS) of stellar matter or a phase transition in the stellar equation of state (EOS).
This would have drastic effects on astrophysical objects and observations, see for example 
\cite{Balkin:2022qer,Balkin:2023xtr,Gomez-Banon:2024oux,Kumamoto:2024wjd,Bartnick:2025scalars}.
Following this train of thought, in this work, we focus on two examples:
a generic light scalar field coupled to electrons and leading to a first-order phase transition, and a lighter version of the QCD axion \citep{Hook:2018jle,DiLuzio:2021pxd,Banerjee:2025zcd}.

We will use the observed white dwarf MRR to
constrain these models of new physics.
See also \cite{Crumpler:2025dvl} for an MRR constraint on linearly coupled ULDM models.
We note that there has been a substantial recent improvement with the MRR observational data. While theoretically, within the SM, this relation is well understood and routinely used in
the analysis of white dwarfs, the empirical evidence has been
problematic until very recently (e.g. \citep{Koester87, Vauclair.Schmidt.ea97, 
Provencal.Shipman.ea98, Joyce.Barstow.ea18, Bedard.Bergeron.ea17}). 
The bulk of the white dwarf population agreed with the theoretical MRR, but the shape of the relation
was almost unconstrained.
This has changed dramatically over the last years due to the use of
very large telescopes for high-quality optical spectra and orbital
elements of eclipsing binaries, well calibrated ultraviolet (UV) spectra from the
\textit{Hubble} telescope, and in particular, excellent distance measurements
from \textit{Gaia} \citep{2023A&A...674A...1G}. This has now resulted in high-precision data for about 30 white
dwarfs, which we use for the comparison with our theoretical models (see Table \ref{tab:MRR_observed}).

In Section \ref{sec:modified_EOS}, we will outline the particle physics models and explain how they lead to a modified EOS.
Section \ref{sec:WDModels} details the calculation of white dwarf models, which lead to a (modified) MRR discussed in Section \ref{sec:ModifiedWD_MR}.
The resulting limits on the particle physics models are shown and discussed in Section \ref{sec:Results}.

\section{The equation of state with new light scalar fields} \label{sec:modified_EOS}
To derive constraints on new scalar fields, we must first understand how they influence the stellar structure of white dwarfs, which was first noted in \cite{Balkin:2022qer} for light QCD axions.
As detailed in the Appendix, we will focus on the so-called negligible-gradient limit where the length scale over which the scalar field changes is much smaller than the size of the star. 
Consequently, the effect of the scalar fields can be fully described by a modification of the white dwarf EOS.
In this section, we will give a brief overview of the particle physics model and modified EOS. More details can be found in App.~\ref{app:particlePhysics} and \cite{Balkin:2022qer,Balkin:2023xtr}.

\begin{figure}
    \centering
    \includegraphics[width=\linewidth]{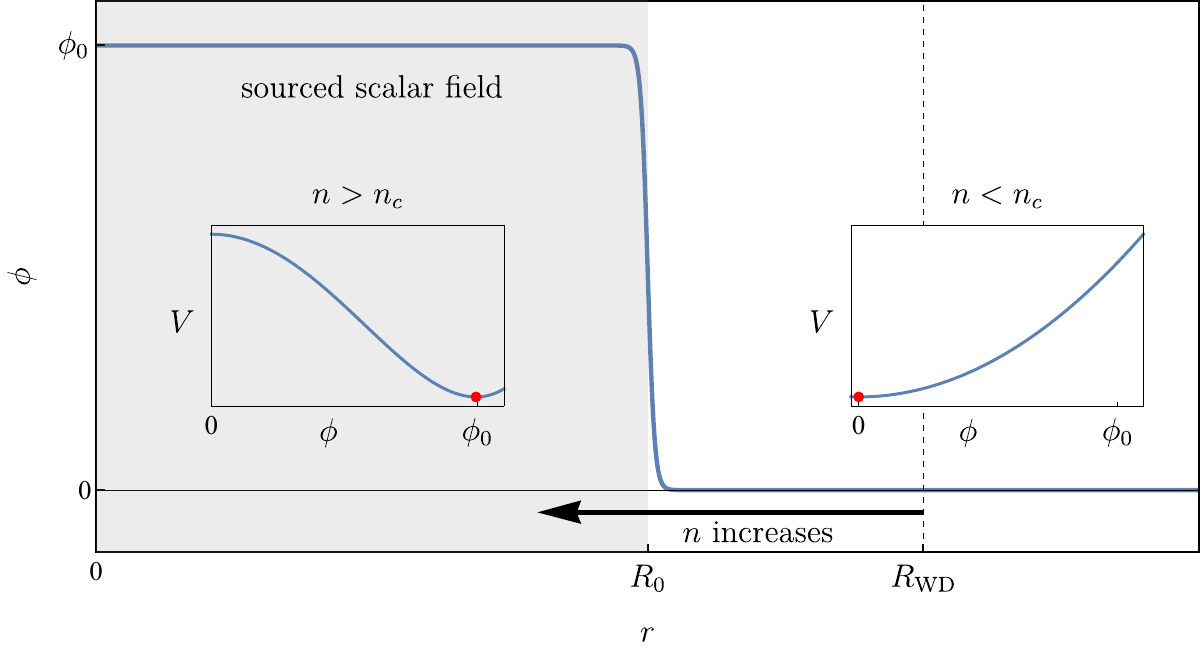}
    \caption{Schematic description of a scalar field profile emerging in a white dwarf.
    The main panel sketches the scalar field as a function of radius $r$ in and around a white dwarf. The dashed line marks the radius of the white dwarf $r=R_\mathrm{WD}$. The insets show the scalar potential at their respective positions. The minimum of the potential, which corresponds to the realized field values, is marked in red. See the text for details.}
    \label{fig:sourcingSketch}
\end{figure}

If a new light scalar field couples quadratically to Standard Model fermions, finite fermion densities influence the scalar potential $V$.
For electron densities $n$ below a critical value $n_c $, $n < n_c$, the scalar potential has a minimum at $\phi=0$ (see right inset in Fig.~\ref{fig:sourcingSketch}).
For the right sign of the coupling, as the density is increased, the scalar mass will decrease and the potential will flatten until the minimum at $\phi=0$ becomes a maximum at $n=n_c$.
Above this critical density, the minimum of the scalar potential is displaced from the origin and lies at a field value $\phi_0\neq0$ (see left inset in Fig.~\ref{fig:sourcingSketch}).
As a consequence, at high densities and neglecting gradients, it is energetically preferred for the scalar field to sit in the new minimum, which we denote by $\phi_0$, and hence develop an expectation value $\expval{\phi} = \phi_0$, similarly to how the Standard Model Higgs field obtains a vacuum expectation value due to its potential being maximized at the origin.

In an analogy to electrodynamics, this is referred to as scalar sourcing:
while a finite charge density sources a classical electric field, a finite number density can source the scalar field; it obtains a classical expectation value.
In a white dwarf, where the density increases from $n \approx 0$ at $r=R_\mathrm{WD}$ as $r$ decreases, the scalar field thus develops a non-trivial profile, interpolating between $\phi(R_\text{WD})=0$ to some $\phi_0(R_0)\neq0$ inside the star with $R_\text{WD}\geq R_0$ (see Fig.~\ref{fig:sourcingSketch}).\footnote{In the case of a new ground state $R_0=R_\text{WD}$, while this work focuses on the case of a first-order phase transition where $R_\text{WD}>R_0$. For details see below and \citep{Balkin:2022qer,Balkin:2023xtr}.}

We work in a weak coupling regime, where no scalar particles are admixed to the white dwarf matter; the whole effect of the scalar field comes from the presence of the expectation value.
This can have an extremely strong backreaction on the white dwarf EOS.
First, let us focus on the scalar field coupled to electrons; we will comment on the axion model below.
The model is described by two free parameters, the scalar mass $m_\phi$ and the coupling to electrons $\dme$, which we choose to be negative $\dme < 0$.
The former fixes the scalar field potential as
\begin{equation}
    V(\phi(x)) = \frac{1}{2} \frac{m_\phi^2 c^2}{\hbar^2} \phi(x)^2 =  \frac{m_e^4 c^4}{\hbar^3} c_m \theta(x)^2,\label{eq:VScalar}
\end{equation}
which we truncate at quadratic order in the scalar field (see \cite{Bartnick:2025scalars} for the importance of higher order terms), and where we defined
\begin{equation}
    \theta(x) = \sqrt{|\dme|\hbar /2}\cdot\phi(x)/\left(M_p c\right) \label{eq:DefTheta}
\end{equation}
and
\begin{equation}
     c_m=m_\phi^2 M_p^2/\left(|\dme| m_e^4\right), \label{eq:DefCm}
\end{equation}
with electron mass $m_e$, where $M_p = \sqrt{\hbar c/8\pi G}$ is the reduced Planck mass.
Here, $c$ is the speed of light and $\hbar$ the reduced Planck constant.
The coupling to the electrons $e$ 
\begin{equation}
    \mathcal{L}_\mathrm{int} = -\frac{\dme}{2 M_p^2 c} m_e \phi(x)^2 \bar{e} e
\end{equation}
can be rewritten as a scalar field-dependent electron mass
\begin{equation}
    m_e(\phi(x)) = m_e \left(1 - \frac{|\dme| \hbar}{2 M_p^2 c^2}\phi(x)^2\right) = m_e \left(1-\theta(x)^2\right).\label{eq:meModScalar}
\end{equation}
This shows that, neglecting scalar field gradients, the theory is described by a single parameter $c_m$, defined in Eq.~\ref{eq:DefCm}.

As described above, the field can be sourced within the white dwarf and take on a local expectation value which follows the minimum of the potential that is fixed by the number density $n$ (or chemical potential $\mu$) and we can set $\phi = \phi(n)$ (or $\phi = \phi(\mu)$) in Eqs.~\eqref{eq:VScalar} and \eqref{eq:meModScalar}.

While the details of finding $\phi(n)$ are outlined in App.~\ref{app:particlePhysics}, here we note the following.
Solving for $\phi(n)$ we find that the field is minimized at $\phi = 0$ for densities $n<n_c$, with the critical density $n_c$, at which the minimum at $\phi=0$ ceases to exist, which is given by
\begin{equation}
    n_c \simeq \frac{m_\phi^2 M_p^2 c^3}{|\dme| m_e \hbar^3} = c_m m_e^3  \frac{c^3}{\hbar^3},\label{eq:nCrit}
\end{equation}
up to relativistic corrections.
At higher densities, $n>n_c$, $\phi$ will have a non-zero value and approach $\theta \to 1$ (where the electrons become massless) at large densities.
Using the effective, density-dependent scalar field values, we can find the EOS including the scalar field.
Besides the scalar-field dependent electron mass, the scalar potential, which behaves like a vacuum energy, has to be included.
To allow for a self-consistent addition of the scalar field effects, we use a simple model for the matter content.
We describe the electrons as a free Fermi gas at finite temperature and combine this with one species of non-relativistic nuclei, following the ideal gas law. 
Consequently, after using charge neutrality, the EOS is given by
\begin{equation}
    \begin{aligned}
         \rho c^2&= \frac{A}{Z} m_N c^2n  + \varepsilon_e(n,T,m_e(\phi(n)))+c V(\phi(n))\\
    p &= p_e(n,T,m_e(\phi(n)))+\frac{n}{Z} k_B T - c V(\phi(n)),
    \end{aligned} \label{eq:EOS_Scalar}
\end{equation}
with nucleon mass $m_N$ and Boltzmann-constant $k_B$.
Here $\rho$, $n$, $\varepsilon_e$, and $p_e$ are the total mass density, electron number density, energy density, and pressure, respectively.
The latter two are given by the corresponding expressions for a free Fermi gas of electrons, where the electron mass has been replaced by the scalar field-dependent mass (Eq.~\eqref{eq:meModScalar}).
$A$ and $Z$ are the nucleon and proton numbers of the nuclei.

\begin{figure}
    \centering
    \includegraphics[width=0.9\linewidth]{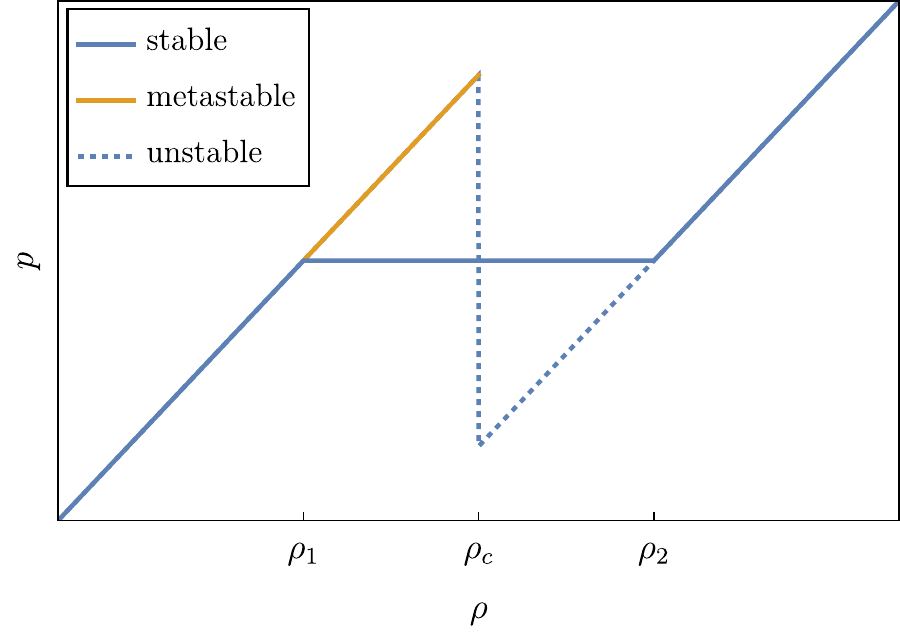}
        
    \includegraphics[width=\linewidth]{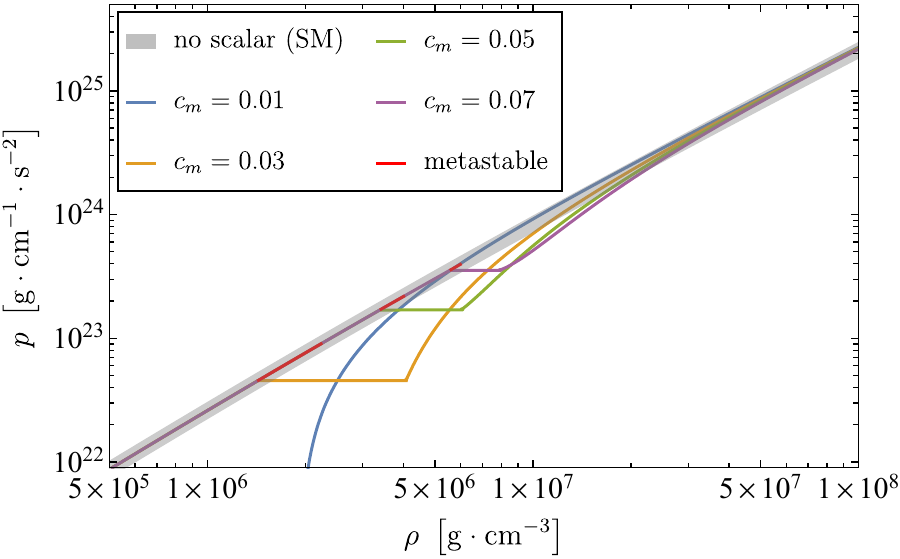}
    \caption{First-order phase transition behaviour in the EOS.
    Top: Schematic sketch of the phase transition behaviour, stable, metastable, and unstable branches, see text.
    Bottom: EOS in the presence of a quadratically coupled scalar field for different values of mass to coupling ratios $c_m$. The EOS without a scalar field is shown in grey and the metastable branch in red. Note that for $c_m=0.01$, the phase transition happens at pressures and densities much lower than visible. The temperature is fixed at $T=10000~\mathrm{K}$ for this example.
    }
    \label{fig:FOPTPlot}
\end{figure}

When including the scalar field, this simple EOS shows a first-order phase-transition behaviour (Fig.~\ref{fig:FOPTPlot}, top).
If we track the scalar field as a function of density up to the critical density, the EOS would follow the blue line first, but then move on to the orange and the blue dashed line at  $\rho=\rho_1$ and $\rho=\rho_c$ respectively.
Here $\rho_c$ is the mass density corresponding to $n_c$, that is $\rho_c=\rho(n_c)$ in Eq.~\eqref{eq:nCrit}.
Up until $\rho=\rho_c$, this EOS corresponds to the SM white dwarf EOS without the inclusion of the scalar field.
However, this EOS is thermodynamically not stable.
Only after performing a Maxwell construction (or, in practice, tracking the scalar field as a function of electron chemical potential), the EOS follows a thermodynamically stable shape marked by the blue solid line in the top panel of Fig.~\ref{fig:FOPTPlot}.
This also defines the mass densities $\rho_1$ and $\rho_2$.
In particular, if the central density of the white dwarf is above the critical density, the stable EOS will describe the entire star.
However, no stable solutions exist with central density $\rho_0$ such that $\rho_c < \rho_0 < \rho_2$.
Similarly, all stellar solutions with $\rho_0 > \rho_2$ will show an abrupt jump from $\rho_2$ to $\rho_1$ as the density decreases with increasing radius, which is typical for first-order phase transitions.

If, instead, the central density is smaller than the critical one ($\rho_0<\rho_c$), in principle, the stable (but sourced) EOS would be energetically favourable (if $\rho>\rho_1$).
This would require a large increase in the scalar field, which is prevented by a potential barrier.
Consequently, solutions with $\rho_0<\rho_c$ track the metastable branch of the EOS, resulting in an ordinary white dwarf.

In our model, including the scalar field quadratically coupled to electrons, we find a first-order phase transition in the EOS for parameters,
\begin{equation}
    0.0094 < c_m < 0.144,
\end{equation}
see also Fig.~\ref{fig:FOPTPlot} (right), where we show the total pressure $p$ as a function of the mass density.
These solutions are shown in various colours in the bottom panel of Fig.~\ref{fig:FOPTPlot} for different values of $c_m$ and are energetically preferred to the SM white dwarf EOS, shown in thick grey. 

For lower values of $c_m<0.0094$, the first-order phase transition behaviour disappears.
Again, there exists a metastable branch for densities $\rho<\rho_c$ with $\phi=0$ and a stable branch which now starts with $p(\rho)=0$ at non-zero densities $\rho = \rho_2$ and $\phi\neq0$. 
The state for $\rho>\rho_2$ with sourced scalar is in a new ground state of matter (NGS) as shown in \cite{Balkin:2022qer}.

Instead, for larger values $c_m>0.144$, the first-order phase transition is replaced by a second-order phase transition.
This is because at large $c_m$ the pressure does not decrease at $\rho_c$, as is the case for the first-order phase transition (FOPT; see top panel of Fig.~\ref{fig:FOPTPlot}). Instead, it still increases, however, more slowly for $\rho>\rho_c$ than for $\rho<\rho_c$, and hence no Maxwell construction is needed. 
The effect ceases to exist with increasing $c_m$ as the slope of the pressure at $\rho>\rho_c$ tends towards the SM white dwarf equation of state.
See also \cite{Bartnick:2025scalars} for a discussion of the phase structure of this model as well as more details on the NGS.\\

At low densities, the scalar field always sits at zero and thus does not affect the white dwarf.
Consequently, we use a more realistic EOS in those regions to better model the envelope and atmosphere, see Sec.~\ref{sec:WDModels}.
Combining the outer EOS with the (modified) free Fermi gas, we obtain the full white dwarf equation of state including the effects of new light scalar fields.
While certainly simplistic, this model captures all the relevant features and as shown below, when omitting the scalar field, well explains the observed mass-radius relation.\\

\begin{figure}
    \centering
\includegraphics[width=\linewidth]{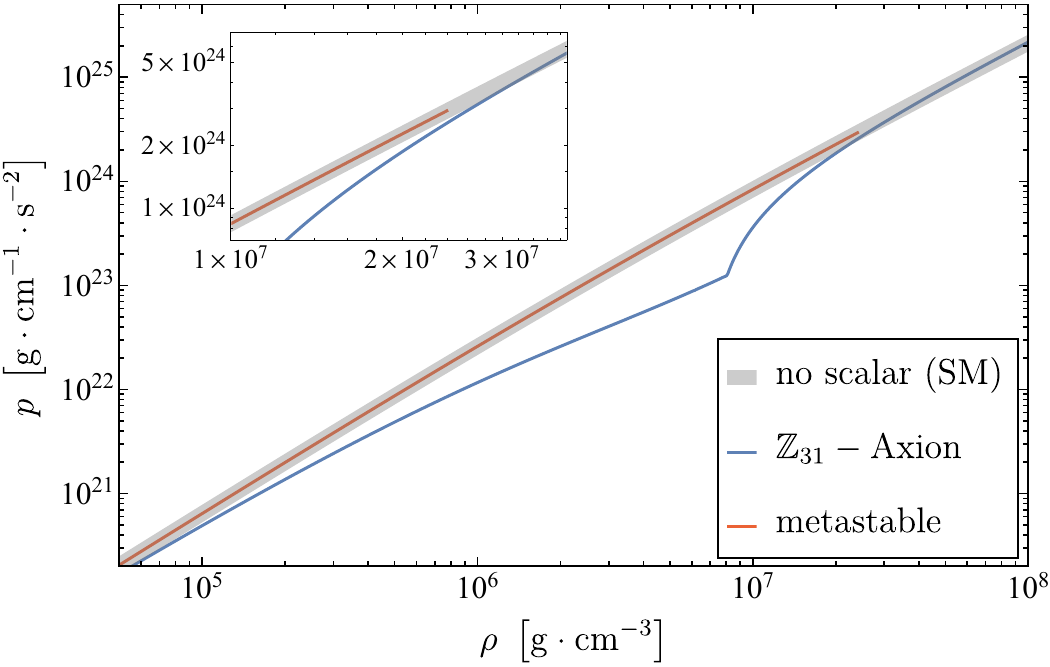}
    \caption{White Dwarf EOS in the presence of a $\mathbb{Z}_{31}$-axion (blue).
    The free Fermi gas EOS without any new physics is shown in grey for comparison. 
    If the central density is below $\rho_c \approx 2.4 \cdot 10^7~\mathrm{g} \cdot \mathrm{cm}^{-3}$, as described in the Appendix, there is a metastable branch (red) tracking the SM EOS.
    The temperature is fixed at $T=10000~\mathrm{K}$ for this example. 
    }
    
    \label{fig:EOS_ZN}
\end{figure}

Instead of coupling to electrons, a new light scalar field might also couple to nucleons.
This will typically lead to a new ground state of matter, resulting in drastic changes of the structure of white dwarfs. The process is therefore tightly constrained  (see, for instance, \citealt{Balkin:2022qer,Balkin:2023xtr}). QCD axion models have such a coupling to nucleons but are a priori not testable in white dwarfs, since $n_c$ is too high. However, axion models, which still solve the strong CP problem but have a lower mass at a fixed decay constant, become testable as first discussed by \citep{Balkin:2022qer}. 
Here we focus on the so-called $\mathbb{Z}_\Ncurl$-model for the lighter than usual QCD-axion developed by \cite{Hook:2018jle}, where the axion mass is suppressed with a factor of roughly $(1/2)^\Ncurl$.
For $\Ncurl = 31$, in particular, this leads to a previously unconstrained cross-over phase-transition, which is different from the NGS, first- and second-order phase transition discussed above, but observable in white dwarfs.
Due to its more complicated vacuum structure, there is also a metastable branch present. We discuss all details in App.~\ref{app:ZnAxion}. Here, we just show the resulting EOS (Fig.~\ref{fig:EOS_ZN}).

\section{Calculation of models for white dwarfs} \label{sec:WDModels}
White dwarfs are the endproduct of the evolution for all stars with
masses up to $\approx 8$ M$_\odot$, that is the vast majority of all
stars. ``Endproduct'' means that the stars have no nuclear energy
production and instead derive their energy output from gravitational
contraction. Because of the extreme densities in the interior ($10^6$
g cm$^{-3}$ and more) the electrons form a degenerate Fermi gas, almost
independent of temperature, which dominates the EOS. In that case,
gravitational contraction does not lead to heating, as in normal
stars, but to a cooling process, which takes billions of years.

The interior structure of white dwarfs, or in general all stars, is
derived from the principles of hydrostatic equilibrium, mass
conservation, energy conservation, and the transport of energy. This
leads to four differential equations, which describe pressure
$P(r)$, mass $m(r)$ inside a sphere of radius $r$, temperature $T(r)$,
and luminosity (total energy flux) $l(r)$. It is equally possible and
often more convenient to use $m$ as independent variable and thus
$r(m)$ as one of the dependent variables.  This is what we also use in our
version; the equations thus are \citep{Koester.Kepler.ea20}

\begin{eqnarray} \frac{dr}{dm} & = &
  \frac{1}{4\pi r^2 \rho}, \\
  \frac{dP}{dm} & = & -\frac{G m}{4 \pi r^4}  \label{eqp},\\
  \frac{d\ln T}{d\ln P} & = &\begin{cases} \frac{3 P \kappa}{64 \pi \sigma T^4 G}\,
  \frac{l}{m}, & \mbox{\qquad (radiative)}\\
  \nabla_{conv}, & \mbox{\qquad (convective)}\end{cases}\\
  \frac{dl}{dm} & = & \epsilon = \frac{L}{M},
  \end{eqnarray}
with total energy output at the surface (luminosity) $L$, total mass $M$, 
mass density $\rho$, $G$ the gravitation constant, $\sigma$ the
radiation constant, $\kappa$ the absorption coefficient, and
$\epsilon$ the energy generation per gram. In the case of white
dwarfs, with no nuclear energy generation, $\epsilon$ is derived from
the gravitational contraction, which can only be calculated with time
dependent evolution. Since we here use only static structures, the
quantity is not available and we use the common approximation
$\epsilon$ = const = $L/M $, which assumes that the energy generation
$\epsilon(m)$ is roughly proportional to $m$. This results in model
structures very close to those from real evolutionary calculations.
The above equations neglect corrections from general relativity, which are negligibly small for the solutions in this work.

If a region in the envelope is unstable for convection, the convective
gradient $\nabla_{conv}$ is calculated with the mixing-length
approximation in the ML2 version \citep{Tassoul.Fontaine.ea90},
assuming a mixing length of 0.8 pressure scale heights.

The parameters used for a specific calculation are the total mass $M$
and total luminosity $L$. For the solution with a Runge-Kutta
integration we need four boundary conditions. These are at the centre
$r(0) = 0, l(0) = 0$, and at the surface of the star $P(M)$ and $T(M)$.
Because of this nature of the boundary conditions, the usual method
to solve the equations is to assume numbers for the missing data at
the centre and at the surface, solve the equations from the surface to
some fitting point, from the centre to this fitting point, and iterate
the guessed boundary values until both integrations agree at the fit
point. In detail, $P(0), T (0)$ and $R(M)$ are guessed and iterated;
from the assumed $R$, and the fixed parameters M and L the effective
temperature and surface gravities can be calculated. These are then
used to obtain $P(M)$ and $T(M)$ from a model of the stellar
atmosphere with these parameters.

For the chemical composition of our models we assume a stellar core of
carbon, surrounded by a helium layer of 1\% of the total mass, and an
outer hydrogen layer of 0.01\%. The transitions between these layers
are calculated from the diffusion equilibrium, which results in smooth
transition layers of a few pressure scale heights thickness. 
This structure is typical of the white dwarfs of spectral type DA
(characterized by Balmer lines in the optical spectra), the most
common type observed. The layered structure is a result of
gravitational separation in the very high gravitational field, in
combination with nuclear burning in the center in previous
evolutionary stages.

In addition to the structure equations we also need an equation of
state, and the opacity of the matter. 
In the outer envelope integration we use the H/He equation of state of \cite{Saumon.Chabrier.ea95},
augmented outside the range of their tables by our own version of the
classical EOS. The latter includes the ideal gas for ions, partially
degenerate electrons, and Coulomb interactions, which are calculated
from numerical fits to Monte Carlo simulations of the One Component
Plasma [\cite{deWitt96}, as cited in \cite{Chabrier.Potekhin98}].

In the interior calculation (95\% of the total mass) different
versions of the EOS with scalar fields or with axions are used (see Sec.~\ref{sec:modified_EOS}).  The
opacities are obtained from the OPAL tables \citep{Iglesias.Rogers96},
and from the Los Alamos opacity
tables \citep{Colgan.Kilcrease.ea16}. Electron conduction data are
from \cite{Potekhin.Baiko.ea99}.

\section{Mass-radius relation of white dwarfs} \label{sec:ModifiedWD_MR}
Numerous studies have discussed the theoretical relation between the
mass and the radius of the white dwarfs, starting from the
zero-temperature models of \cite{Chandrasekhar35, Chandrasekhar39,
  Hamada.Salpeter61} to the latest calculations including finite
temperature effects of the Montreal
\footnote{\url{https://www.astro.umontreal.ca//~bergeron/CoolingModels/}}
and La Plata (Buenos Aires)
\footnote{\url{https://evolgroup.fcaglp.unlp.edu.ar/modelos.html}} groups
\citep[e.g.][]{Bedard.Bergeron.ea17, Panei.Althaus.ea00}.

\begin{figure}
  \centering
  \includegraphics[angle=0,width=\linewidth]{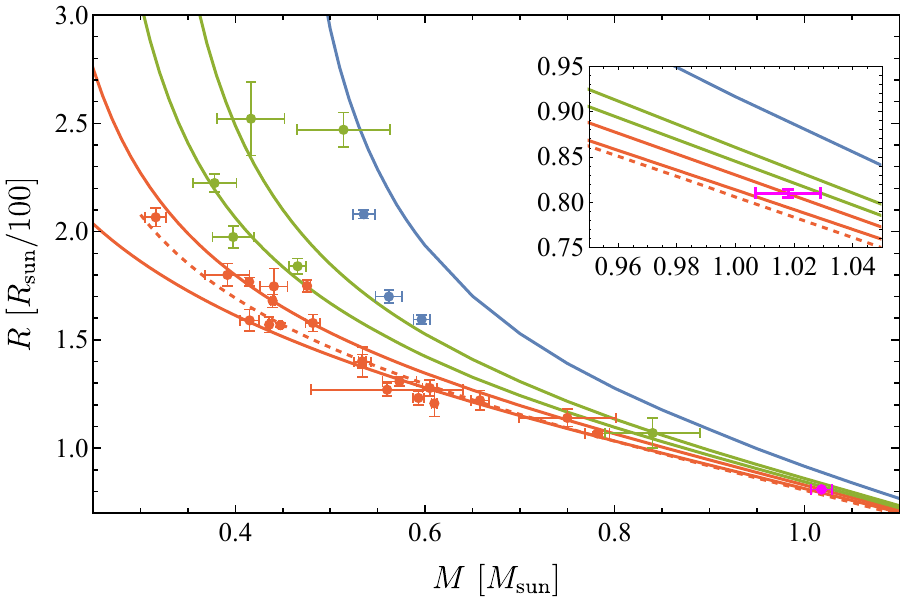}
  \caption{Mass-radius relations with the classical SM equation of state for
    the effective temperatures $5000, 15\,000, 25\,000, 35\,000, 60\,000\,\mathrm{K}$. Colour
    codes different temperature ranges; red: $5000-20000\,\mathrm{K}$, green:
    $20000-40000\,\mathrm{K}$, blue: $>40000\,\mathrm{K}$.  The crosses are the observed
    masses and radii described in the text. They use the same colour
    coding as the theoretical relations, except for Sirius B, which we plot in magenta for better visibility.
    The dashed line is the $15000\,\mathrm{K}$ relation from the LaPlata group. 
    The inset shows an enlarged plot around Sirius B. 
    }\label{mrrCHeHSM}
\end{figure}

\begin{figure}
  \centering
  \includegraphics[angle=0,width=\linewidth]{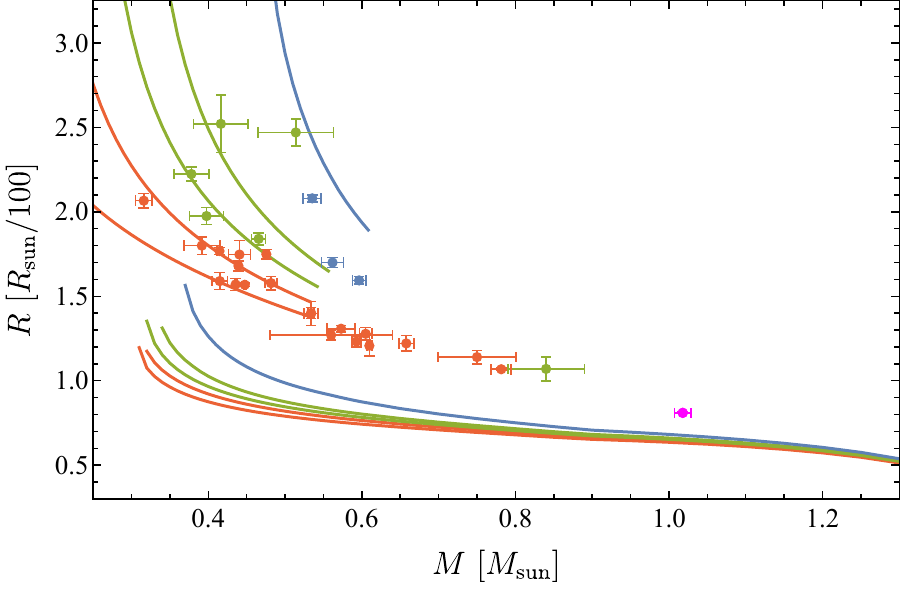}
  \caption{Mass-radius relations with scalar EOS for the
    effective temperatures $5000, 15000, 25000, 35000$ and $60000\,\mathrm{K}$ (from bottom) and $c_m =0.03$.
    The five upper curves are stable branches for low masses, extended by metastable branches depending on the central density (see Fig.~\ref{fig:FOPTPlot}).
    The crosses are the observed masses and radii as in
    Fig.~\ref{mrrCHeHSM}, which the theory with the new scalar field badly fails to explain.
    Colouring of objects and mass-radius relations as in Fig.~\ref{mrrCHeHSM}.
    }
    \label{scalar03}
\end{figure}

\begin{figure}
  \centering
  \includegraphics[width=\linewidth]{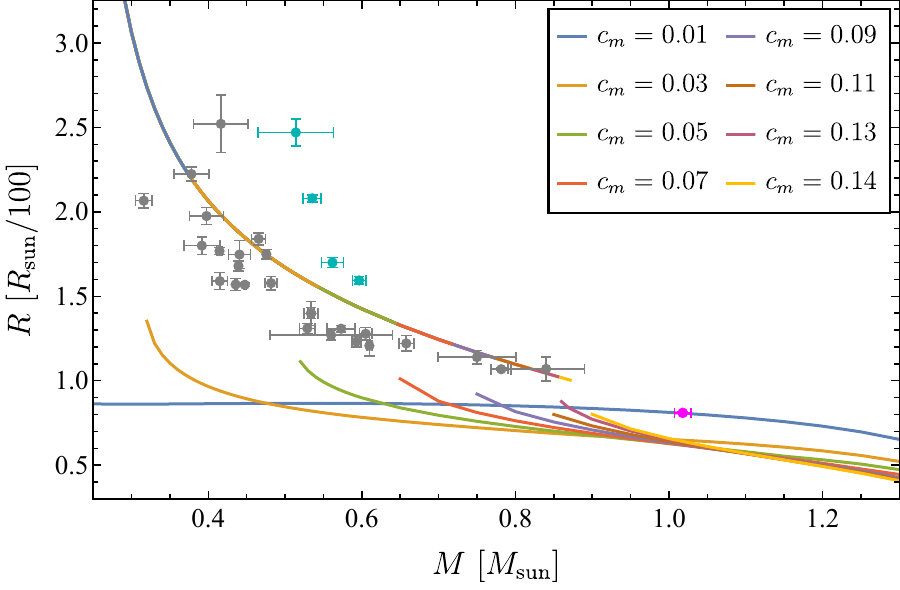}
  \caption{Dependence of the theoretical mass-radius relation for $T_\mathrm{eff}$ =  $25000\,\mathrm{K}$ on the parameter $c_m = 0.01 - 0.14$.
    The objects in cyan have effective temperatures above 35000\,K, Sirius B ($T_\mathrm{eff} = 25369\pm63\,\mathrm{K}$) is again shown in magenta. Note that the model with $c_m=0.01$ is very strongly first-order and thus shows the strongest deviation (see Fig.~\ref{fig:FOPTPlot}).
    }\label{mrrc}
\end{figure}

\begin{figure}
  \centering
  \includegraphics[width=\linewidth]{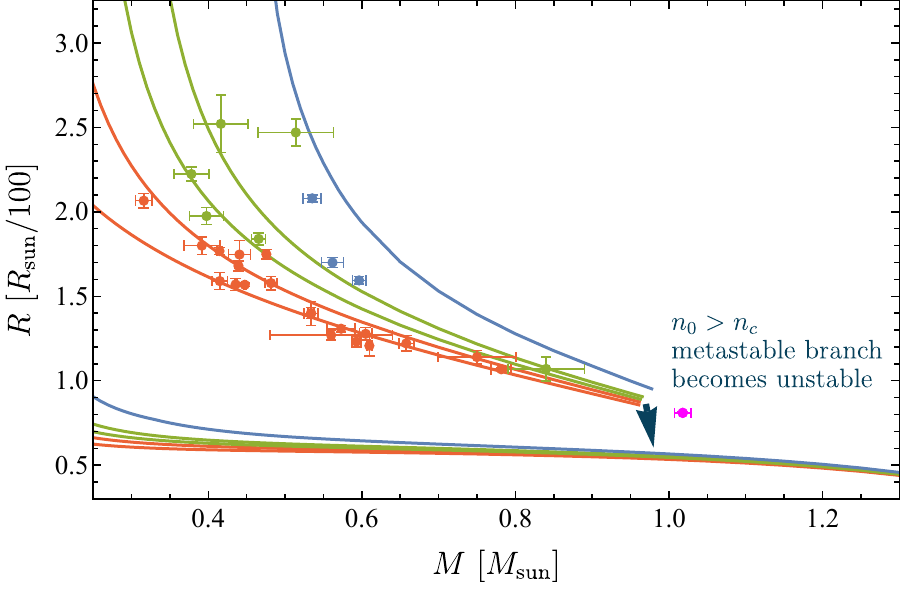}
  \caption{ Mass-radius relations with $\Ncurl = 31, \mathbb{Z}_\Ncurl$-axion modified EOS. The MRRs from the stable (lower curves) and metastable (upper curves) branches of the EOS are shown for the effective temperatures $5000, 15\,000, 25\,000, 35\,000$, and $60\,000\,\mathrm{K}$ (from the bottom). The metastable branches end, when the central
    density $n_0$ equals the critical one $n_c$.  Increasing the
    central density beyond this point would trigger the phase
    transition in the star and lead to an unstable collapse, as
    denoted by the dark blue arrow. The metastable branch can
    explain nearly all well-observed white dwarfs, Sirius B, thought (shown in
    magenta), is in clear contradiction to the prediction of this model.
    Colouring of MRRs and objects based on temperature as in Fig.~\ref{mrrCHeHSM}.
  }\label{fig>axionMrr}
\end{figure}

The dependence of the theoretical relation (MRR) on mass, luminosity,
effective temperature, interior, and outer chemical composition is well
understood. The relation is routinely used to determine for example masses
from spectroscopically determined surface gravities.  The gravity
$g=GM/R^2$ provides a second relation between mass and radius, such
that both quantities can be determined.

On the other hand, the observational confirmation of the MRR through
comparison with observations has until recently met with only limited
success. Various methods have been used:
\begin{itemize}
\item radius from surface gravity g + dynamical mass from white dwarfs
  in binary systems,
\item radius and mass from total luminosity L + distance + effective
  temperature + surface gravity,
\item radius and mass from surface gravity + gravitational redshift
  $v_{rs} = G M/R c$. This is only
  possible, when the space velocity of the white dwarf is known from a
  companion star with negligible $v_{rs}$.
\item radius and mass from total luminosity L + distance + effective
  temperature + gravitational redshift,
\item radius and mass from eclipsing white dwarfs in binary systems +
  gravitational redshift.
\end{itemize}

Most of the methods involve an analysis of the spectra through
comparison with theoretical models of the atmospheres to obtain
effective temperatures and surface gravities. This typically involves
errors of 3-5\% for $T_{\text{eff}}$ and 15-20\% for $g$. The distances --
before the data from the astrometric satellite \textit{Gaia} became available --
could easily add another 20-30\% error. Another problem is that the
white dwarfs are strongly concentrated around an average mass of 0.6
solar masses and very few are near the low and high ends of the mass
distribution. While the bulk of the MRR distribution did agree with the
theoretical prediction, the shape was not empirically confirmed.  All
this contributed to the situation, that while the number of known
white dwarfs had increased from 3 in 1920 to 359000 in 2021 \citep{GentileFusillo.Tremblay.ea21}, there were
until recently only a handful of objects with significantly smaller
errors than described above.

The situation changed dramatically with the data from eclipsing
binaries with white dwarfs together with gravitational redshifts
presented by \cite{Parsons.Gaensicke.ea17}. This method is not only very
accurate, but also independent of theoretical model atmospheres. For
the first time a larger sample with typical errors of 2\% for mass and
radius became available.

We use this observational sample of 26 objects together with the
only 5 objects with comparable accuracy known previously: the 3
classical objects Sirius~B \citep{Joyce.Barstow.ea18}, Procyon~B
\citep{Bond.Gilliland.ea15}, 40~Eri~B \citep{Bond.Bergeron.ea17}, and
new results from microlensing white dwarfs Stein~2051B
\citep{Sahu.Anderson.ea17}, and LAWD~37 \citep{McGill.Anderson.ea23}. 
The masses, radii and effective temperatures of these objects are given in Table \ref{tab:MRR_observed}.

\begin{table*}
    \centering
    \caption{Observed data for the white dwarf mass radius relation}
    \begin{threeparttable}
    \begin{tabular}{lllll}
    \hline
    $R~[R_\mathrm{sun}]$ & $M~[M_\mathrm{sun}]$ & $T_\mathrm{eff}~[\mathrm{K}]$ & Source\tnote{\tnotelink} & Name\\
    \hline
$0.01749\pm 0.00028$ & $0.4756\pm 0.0036$ & $17838\pm 482$ & a \\
$0.02521\pm 0.0017$ & $0.4164\pm 0.0356$ & $29969\pm 679$ & a \\
$0.01221\pm 0.00046$ & $0.6579\pm 0.0097$ & $15909\pm 285$ & a \\
$0.01578\pm 0.00039$ & $0.4817\pm 0.0077$ & $14901\pm 731$ & a \\
$0.02224\pm 0.00041$ & $0.378\pm 0.023$ & $22497\pm 60$ & a \\
$0.02066\pm 0.00042$ & $0.316\pm 0.011$ & $11864\pm 281$ & a \\
$0.017\pm 0.0003$ & $0.5618\pm 0.0142$ & $50000\pm 673$ & a \\
$0.0208\pm 0.0002$ & $0.5354\pm 0.0117$ & $63000\pm 3000$ & a \\
$0.01068\pm 0.00007$ & $0.7816\pm 0.013$ & $14220\pm 350$ & a \\
$0.01568\pm 0.00009$ & $0.4475\pm 0.0015$ & $7540\pm 175$ & a \\
$0.01398\pm 0.0007$ & $0.534\pm 0.009$ & $8272\pm 580$ & a \\
$0.01747\pm 0.00083$ & $0.4406\pm 0.0144$ & $13957\pm 531$ & a \\
$0.0184\pm 0.00036$ & $0.4656\pm 0.0091$ & $24569\pm 385$ & a \\
$0.0131\pm 0.0003$ & $0.529\pm 0.01$ & $3570\pm 100$ & a \\
$0.01594\pm 0.00022$ & $0.5964\pm 0.0088$ & $46783\pm 7706$ & a \\
$0.0247\pm 0.0008$ & $0.514\pm 0.049$ & $37400\pm 400$ & a \\
$0.01401\pm 0.00032$ & $0.5338\pm 0.0038$ & $10644\pm 1721$ & a \\
$0.01768\pm 0.0002$ & $0.4146\pm 0.0036$ & $12221\pm 765$ & a \\
$0.01278\pm 0.00037$ & $0.605\pm 0.0079$ & $10210\pm 87$ & a \\
$0.0159\pm 0.0005$ & $0.415\pm 0.01$ & $6000\pm 200$ & a \\
$0.0168\pm 0.0003$ & $0.4393\pm 0.0022$ & $17707\pm 35$ & a \\
$0.01207\pm 0.00061$ & $0.6098\pm 0.0031$ & $8500\pm 500$ & a \\
$0.018\pm 0.00052$ & $0.3916\pm 0.0234$ & $12491\pm 312$ & a \\
$0.01975\pm 0.0005$ & $0.3977\pm 0.022$ & $20837\pm 773$ & a \\
$0.0107\pm 0.0007$ & $0.84\pm 0.05$ & $34500\pm 1000$ & a \\
$0.0157\pm 0.00036$ & $0.4356\pm 0.0016$ & $7740\pm 73$ & a \\
$0.0127\pm 0.0003$ & $0.56\pm 0.08$ & $7837\pm 83$ & b & LAWD37 \\
$0.01308\pm 0.0002$ & $0.573\pm 0.018$ & $17200\pm 110$ & c & 40 Eri B \\
$0.008098\pm 0.000046$ & $1.018\pm 0.011$ & $25369\pm 63$ & d & Sirius B \\
$0.01232\pm 0.00032$ & $0.593\pm 0.006$ & $7740\pm 50$ & e & Procyon B \\
$0.0114\pm 0.0004$ & $0.75\pm 0.051$ & $7122\pm 181$ & f & Stein 2051 B \\
    \hline
    \end{tabular}
    \begin{tablenotes}
         \item[\tablenotelabel] a: \cite{Parsons.Gaensicke.ea17}, b: \cite{McGill.Anderson.ea23}, c: \cite{Bond.Bergeron.ea17}, d: \cite{Bond:2017Sirius}, e: \cite{Bond.Gilliland.ea15} and f: \cite{Sahu.Anderson.ea17}
    \end{tablenotes}
    \label{tab:MRR_observed}
     \end{threeparttable}
\end{table*}
  
As a first test of our approach, we calculate the MRR using the classical
EOS described in Section~\ref{sec:WDModels} for masses between approximately
0.2 and 1.3 M$_\odot$ for effective temperatures 5000, 15\,000, 25\,000,
35\,000, and 60\,000~K, which covers the entire range of the observational
sample.

Fig.~\ref{mrrCHeHSM} shows these theoretical relations for a wide
range of effective temperatures. The comparison with the relation for
15000\,K from the website of the LaPlata group shows a reasonable
agreement, considering the simplified EOS used for the interior in our
calculations. The comparison with the best available observed data
also shows a good agreement between theory and observation.

In the next step, we have calculated white dwarf models and mass-radius
relations with different versions of our non-classical
EOS leading to first-order phase transitions. Fig.~\ref{scalar03} shows the result for the scalar EOS with
coupling constant $c_m = 0.03$. Obviously, the models fail badly to reproduce the observed MRR. The metastable models lead to better results but their branches stop at $\sim$0.55 M$_{\odot}$.

Fig.~\ref{mrrc} investigates the dependence of the non-classical FOPT EOS on the coupling parameter $c_m$. It shows the MRR at
$T_\mathrm{eff}$ = 25\,000\,K with the parameter $c_m$ in the range 0.01 to 0.14. All models with a phase transition fail badly to reproduce the observed radii. 
For $c_m \gtrsim 0.03$, the sourced branch of the EOS is in tension with all considered white dwarfs.
However, the unsourced and metastable branches of the EOS come closer to the observations.
Increasing the value of $c_m$ increases the critical density (see Eq.~\eqref{eq:nCrit} and Fig.~\ref{fig:FOPTPlot}) and thus the fraction of white dwarfs that can be explained even in the presence of a scalar field.
But, very obviously, even the model with the largest possible $c_m$, $c_m = 0.14$, cannot reproduce the radius of Sirius B.
On the other hand, models with smaller $c_m$ around $c_m \approx 0.01$ fit Sirius B nicely, but do not reproduce the radii of the less massive objects.
This strong deviation with respect to the other models can be understood from the EOS (compare Fig.~\ref{fig:FOPTPlot}, bottom), which shows a corresponding strong deviation through a large part of the white dwarf stellar interior.
At this small value of $c_m$, the phase transition is very strongly first-order since it is very close to the case of an NGS.
To sum up, the MRR provides a crucial constraint on the nature of these particles and rules out the complete range of coupling parameters leading to an FOPT.

In a final step, we test the $\mathbb{Z}_\Ncurl, \Ncurl = 31,$ QCD-axion EOS which leads to a cross-over phase transition. The result is shown in Fig.~\ref{fig>axionMrr}. Again, the metastable branch is able to match most of the observed data. However, it fails to reproduce the mass and radius of Sirius B. Thus, based on this significant disagreement with the observed radius at the high mass of Sirius B, we can rule out this model as well.

\section{Results and Conclusions} \label{sec:Results}
For the scalar field coupling to electron the comparison with the observed white dwarf mass radius data has provided a very strong constraint. All models that lead to a first-order phase transition can be excluded. The part of the theoretical MRRs corresponding to the metastable branch of the EOS can reproduce some of the observed data at lower masses, in particular, when $c_m$ (or the critical density $\rho_c$) is increased. However, for the highest possible value leading to an FOPT $c_m$ = 0.14, the resulting mass and radius is in clear disagreement with the observed values of Sirius B. Given the precision of the determination of the mass and radius of Sirius B, this is a very clear exclusion. 

\begin{figure}
    \centering
    \includegraphics[width=\linewidth]{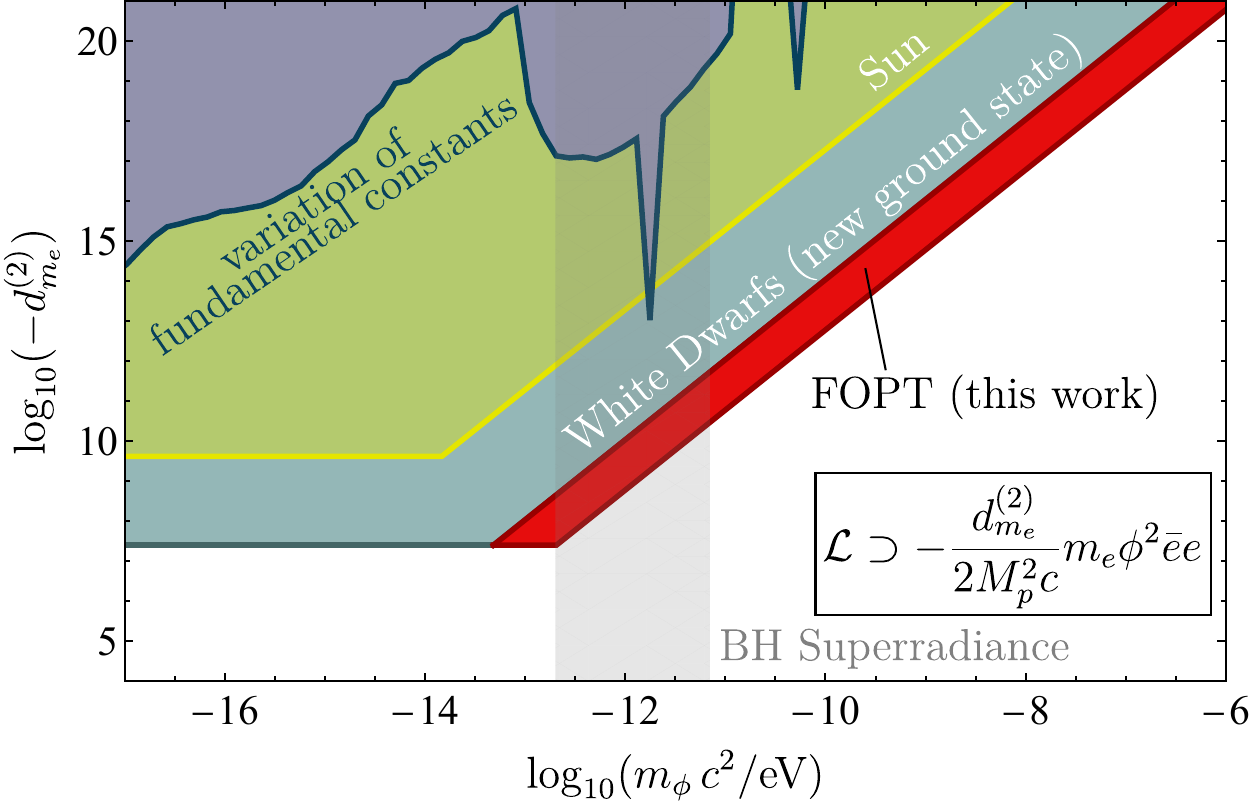}
    \caption{Constraint on scalars-fields quadratically coupled to electrons in the plane of scalar mass $m_\phi$ and coupling to electrons $\dme$. 
    Constrained on the scalar field due to the induction of a first-order phase transition derived in this work are shown in red. 
    For a discussion of other limits, see the main text.
    }
    \label{fig:exclusionPlotScalar}
\end{figure}

To compare this new result with previous work we use the definition of $c_m$ (Eq.~\eqref{eq:DefCm}) and translate back to the typical parameter space of scalar-particle mass $m_\phi$ and coupling parameter $\dme$. We then obtain the red exclusion area shown in Fig.~\ref{fig:exclusionPlotScalar}. The left and right diagonal borderlines of this area correspond to $c_m$ = 0.0093 and 0.144, respectively.
For small coupling parameters $-\dme \le$ 2.5 $\times 10^7$ (independent of the scalar mass $m_\phi$), gradient effects become important.
They prevent significant field displacements in a white dwarf, the scalar field is no longer sourced, and the EOS and predicted MR-curves return to the one of the Standard Model, ending the excluded region towards the bottom.

Above and to the left of the red area, $c_m$ becomes lower.
At $c_m \lesssim 0.0093$, the scalar field no longer leads to a first-order phase transition but a new ground state in the white dwarf EOS.
The corresponding blue-grey area in Fig.~\ref{fig:exclusionPlotScalar} has been investigated by \cite{Balkin:2022qer,Balkin:2023xtr} and 
\cite{Bartnick:2025scalars}: the new ground state EOS leads to a complete change of the structure of white dwarfs and to forbidden gaps in their mass radius relation in clear disagreement with the observations. 

For lower critical densities, the new ground state would also occur in the Sun. Through a considerably lower electron mass in the solar atmosphere, this would lead to solar spectra incompatible with observations \citep{Bartnick:2025scalars}. This area is coloured yellow. The grey exclusion area results from the observation of rapidly rotating solar mass black holes, which would not exist with light scalars in this parameter domain as they extract energy from the black hole through "superradiance" leading to a spin-down \citep{Brito:2015oca, Witte:2024drg}. The blue exclusion area results from precision experiments on Earth using quantum clocks and interferometry \citep{Kennedy:2020bac,Oswald:2021vtc,Vermeulen:2021epa,Savalle:2020vgz,Aharony:2019iad,Antypas:2019qji,Aiello:2021wlp,Branca:2016rez,Sherrill:2023zah,Kobayashi:2022vsf,Campbell:2020fvq,Zhang:2022ewz,Oswald:2025bih}.
Note that these experimental bounds only apply if the scalar field accounts for a large fraction of the dark matter abundance.
Here, we translate the bounds directly from the linear case, assuming the background dark matter density.
There might be changes due to finite density effects leading to non-trivial profiles around the Earth or the Sun (similar to sourcing, but without the scalar mass becoming tachyonic), see \citep{Hees:2018fpg,Banerjee:2022sqg,Budker:2023sex,Banerjee:2025dlo,delCastillo:2025rbr,Grossman:2025cov,Gan:2025nlu}.
\\

\begin{figure}
    \centering
    \includegraphics[width=\linewidth]{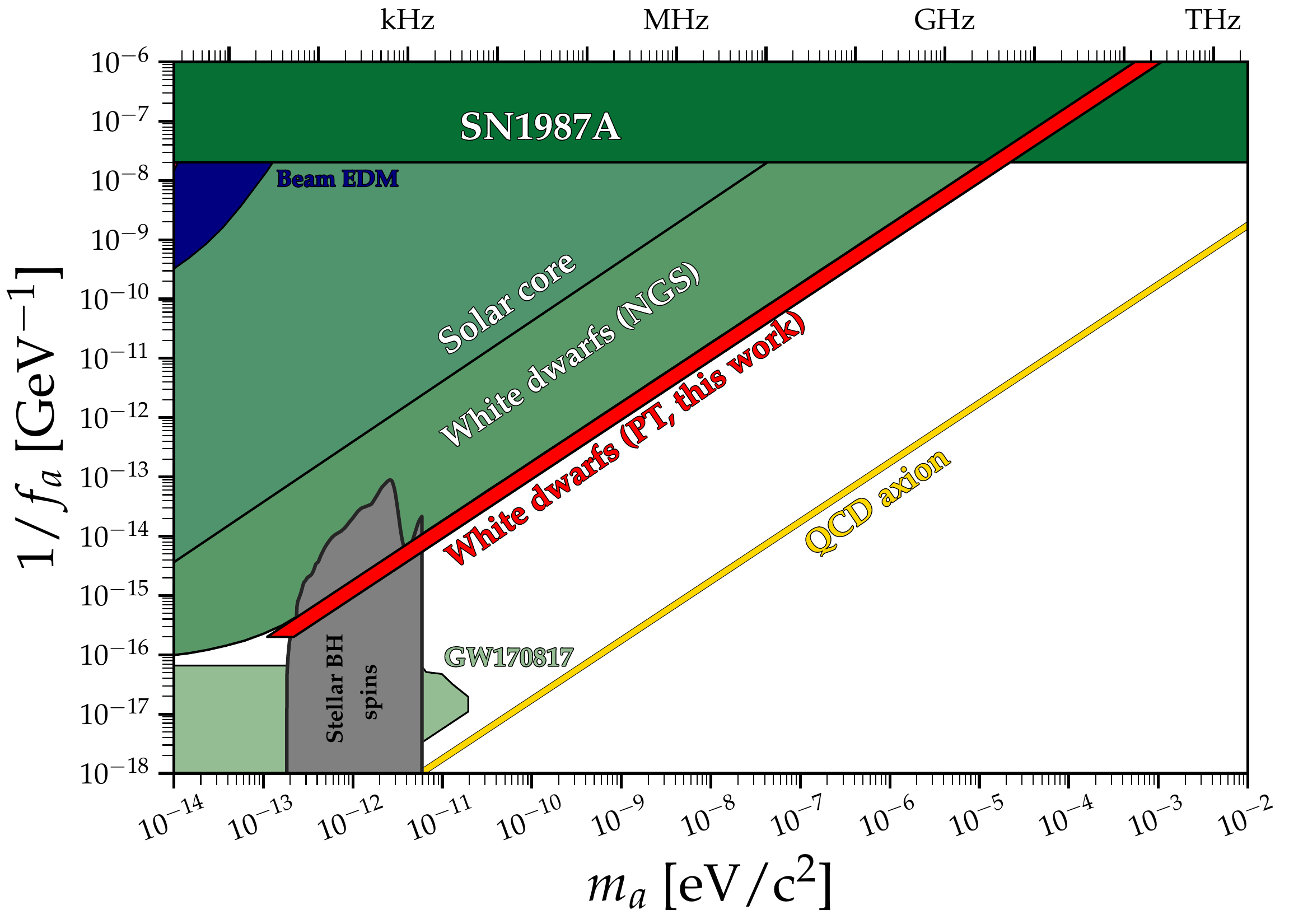}
    \caption{Constraints on $\mathbb{Z}_\Ncurl$ axion models as a function of axion mass $m_a$ and decay constant $f_a$. Constraints due to a (modified) crossover phase transition in white dwarfs derived in this work are shown in red. The $m_a-f_a$ relation in the case $\Ncurl=1$, the standard QCD axion, is given by the yellow line; other constraints are discussed in the main text. Plot modified from \citet{AxionLimits}.}
    \label{fig:exclusionPlotZn}
\end{figure}

In a similar way as for the light scalar field coupling to electrons, we can also discuss the constraints obtained for the $\mathbb{Z}_\Ncurl$-axion. Using the potential as defined in the Appendix in Eq.~\eqref{eq:ZnPotential}, we translate our exclusion into the space of axion mass $m_a$ and coupling constant $f_a$. This is done in Fig.~\ref{fig:exclusionPlotZn}. The red exclusion area is obtained for the case of an EOS corresponding to $\Ncurl=31$ and the comparison with the observed white dwarf MRR carried out in this work. 

The white area below the new red exclusion area corresponds to smaller $\Ncurl$. Here, phase transitions occur in a similar way as for $\Ncurl=31$ but the corresponding metastable branch is able to reproduce the radius of Sirius and no exclusion is obtained. 
As in the scalar case, if the decay constant is too large (i.e.~weak coupling limit), gradient effects prevent axion sourcing and thus the crossover phase transition in white dwarfs, which ends the exclusion for large $f_a$.
In principle, the precise shape of the exclusion at large $f_a$ requires solving the full coupled equations describing the scalar field and the white dwarf, which has been done in \citep{Balkin:2022qer,Balkin:2023xtr} but goes beyond the scope of this work.
Consequently, as in the scalar case discussed above, we set the limit by estimating the typical scale of the axion gradient and comparing it to the white dwarf size.
To be conservative, we end the exclusion slightly earlier compared to \cite{Balkin:2022qer}, where full gradient effects are computed. 

The exclusion areas at higher coupling constants (large $\Ncurl$) results from a NGS in white dwarfs \citep{Balkin:2022qer} and sourcing in the solar core \citep{Hook:2017psm}. 
The bottom green area is also related to axion sourcing, in particular the influence it would have on the neutron star merger GW170817 \citep{Zhang:2021mks}.

Again, black hole superradiance \citep{Witte:2024drg} leads to the exclusion shown in grey. 
Note that, for the $\mathbb{Z}_\Ncurl$-axion, the quartic-self interactions $\lambda a^4$ are enhanced by a factor $\Ncurl^2$ at fixed mass and decay constant compared to the normal QCD-axion case; consequently, the superradiance bounds are correspondingly weaker.
Axion cooling leads to the constraints obtained from supernova 1987A \citep{Springmann:2024ret}.
Finally, experiments such as Beam EDM \citep{Schulthess:2022pbp}
lead to constraints at very low axion mass, if an axion dark matter background is assumed \citep{Banerjee:2025dlo,delCastillo:2025rbr}.
We finally note that sourcing in neutron stars can also constrain light axions due to anomalously fast cooling \citep{Gomez-Banon:2024oux,Kumamoto:2024wjd}.
Since these limits rely on the presence of a new ground state for which heat-blanketing envelopes can be absent, they are fundamentally different from traditional cooling bounds (such as \cite{Raffelt:1996wa,Buschmann:2021juv,Springmann:2024mjp}). Importantly, the $\mathbb{Z}_\Ncurl$-potential has no new ground state at neutron star densities \citep{Balkin:2022qer} and hence avoids these limits, which is why they are omitted in Fig.~\ref{fig:exclusionPlotZn}.
\\

In summary, Fig.~\ref{fig:exclusionPlotScalar} and Fig.~\ref{fig:exclusionPlotZn} demonstrate the power of astrophysical observations in constraining particle physics models. White dwarfs and their mass-radius relation can be used to probe an exciting part of the parameter space for scalar fields that couple quadratically to the Standard Model and turn out to be highly competitive with dedicated experimental searches. In both the case of a scalar coupling to electrons as well as a $\mathbb{Z}_\Ncurl$ axion, white dwarfs probe a large portion of otherwise unconstrained parameter space. Over a large range of masses, the lowest coupling excluded at each mass is set by the incompatibility of phase transitions with the white dwarf mass-radius relation as derived in this paper.
Although the EOS signatures of such phase transitions are subtle, the precision of modern white-dwarf MRR predictions and measurements, especially for Sirius B, now allows us to probe them with confidence.
\\

\section*{Acknowledgements}
    The authors (KB, RK, AW) acknowledge support by the Munich Excellence Cluster Origins and the Munich Institute for Astro-, Particle and BioPhysics (MIAPbP) funded by the Deutsche Forschungsgemeinschaft (DFG, German Research Foundation) under Germany's Excellence Strategy – EXC-2094 – 390783311. KB and AW are also supported by the Collaborative Research Center SFB1258. SS is supported by the Swiss National Science Foundation under contract 200020-213104 and acknowledges the hospitality of the CERN theory group.
    KS is supported by a research grant from Mr. and Mrs. George Zbeda and by the Minerva Foundation.

\section*{Data Availability}
Data on the EOS and MRR are available from the authors upon reasonable request.

\bibliographystyle{mnras}
\bibliography{wdnew,wdScalars}

\appendix

\section{Some particle physics details} \label{app:particlePhysics}
In the following, we will present additional details on the sourcing of the scalar field quadratically coupled to electrons (\ref{app:scalarElectrons}) as well as the $\mathbb{Z}_\Ncurl$-axion (\ref{app:ZnAxion}). 
\subsection{Scalar field coupling to electrons} \label{app:scalarElectrons}
Describing a new scalar field $\phi$ coupling to electrons $e$ we consider the Lagrangian
\begin{equation}
	\mathcal{L} = \bar{e}\left(i\slashed{\partial}-\frac{m_e c}{\hbar}\right)e + \frac{1}{2} \partial_\mu \phi \partial^\mu \phi- \frac{m_\phi^2c^2}{2\hbar^2}  \phi^2 - \frac{\dme}{2 M_p^2 c} m_e \phi^2 \bar{e}e . \label{eq:ScalarElectronLagrangian}
\end{equation}
Here, we only include quadratic couplings between the scalar field and the fermions.
Since this Appendix focuses on particle physics details, we will use natural units where $\hbar=c=1$ from now on.
In general, linear couplings would also be present, but they are tightly constrained experimentally, e.g., from fifth force searches \citep[see e.g.][]{Schlamminger:2007ht,Lee:2020zjt,Tan:2020vpf}.
By imposing a discrete symmetry $\mathbb{Z}_2$ symmetry $\phi \to - \phi$, linear couplings are forbidden, making the leading interactions quadratic in $\phi$.
In general, $\dme$ could have any sign; here, we focus on the negative case $\dme<0$, which has profound implications on stellar structure.

In a leading approximation, a finite electron number density $n_e$ can be understood as an expectation value, $\expval{\bar{e} e} \approx n_e$. 
Consequently, at finite density, the scalar field effective mass is given by
\begin{equation}
    m_{\phi,\mathrm{eff}}^2 = m_\phi^2 - \frac{|\dme|}{M_p^2} m_e n_e.
\end{equation}
In the case of negative coupling, for number densities $n_e$ above a critical density $n_e^c$
\begin{equation}
    n_e^c \simeq\frac{1}{|\dme|} \frac{m_\phi^2 M_p^2}{m_e}
\end{equation}
the scalar mass becomes tachyonic $m_{\phi,\mathrm{eff}}^2<0$.
Consequently, the vacuum scalar-field value $\phi=0$ is no longer stable, and the field develops a large, classical expectation value in regions of dense matter; it gets sourced.

Deriving the full in-density scalar equations of motion, one finds a coupled system together with the equations of stellar structure, describing stars with sourced scalar fields, see \cite{Balkin:2022qer,Balkin:2023xtr} for the limit of negligible temperature.
Solving this system in full generality goes beyond the scope of this work.
In most of the parameter space a simplifying limit exists.
Gradients of the scalar field $\phi$ have an associated energy cost, and the field will change on scales of the order of
\begin{equation}
    \lambda_\phi \approx M_p \sqrt{\frac{2}{m_e |\dme| n_e }}  = 26~\mathrm{km}~\left(\frac{10^{10}}{|\dme|}\right)^{1/2} \left(\frac{m_e^3}{n_e}\right)^{1/2},
\end{equation}
where $n_e$ is given by typical white dwarf electron number densities.
If $\lambda_\phi \gg R_\mathrm{WD}$, the scalar field will not be sourced, and the white dwarf will not be affected by its presence.
In contrast, for $\lambda_\phi \ll R_\mathrm{WD}$, we can neglect all field gradients, and the scalar field will simply track the minimum of the in-density effective potential, which allows us to rewrite the theory in terms of an effective EOS.
In the following, we will focus on this limit.

To obtain the effective potential, we notice from the Lagrangian, eq.~\eqref{eq:ScalarElectronLagrangian}, that at any given expectation value for the scalar field $\phi$, the electron mass depends on that value as
\begin{equation}
	m_e(\phi) = m_e \left(1 - \frac{|\dme|}{2 M_p^2}\phi^2\right).
\end{equation}
This allows a description of the electron effects on the scalar field in terms of an effective potential.
The finite fermion density contribution to the effective potential is given by the corresponding fermion thermodynamic potential with the mass replaced by the scalar-dependent mass.
To correctly model the thermodynamics of the phase transition, it is easiest to work in the grand canonical ensemble, where one finds
\begin{equation}
    V_\mathrm{eff}(\phi, \mu) = V(\phi) + j_e(\mu,m_e(\phi) =  \frac{1}{2} m_\phi^2 \phi^2  - p_e\left(\mu,m_e(\phi)\right), 
\end{equation}
since the grand canonical potential is $j_e = - p_e$.
Here $\mu$ is the chemical potential of electrons.
Since we are considering changing fermion masses, the rest mass energy of the fermions must be included in the chemical potential.
Solving the scalar equation of motion, neglecting gradient terms,
\begin{equation}
    \dv{V_\mathrm{eff}(\phi,\mu)}{\phi} = 0,
\end{equation}
i.e. minimizing $V_\mathrm{eff}(\phi)$ at a given $\mu$ yields $\phi_\mathrm{eq}=\phi(\mu)$ the equilibrium expectation values of the scalar field.
This directly gives $m_e^*(\mu) = m_e(\phi_\mathrm{eq}(\mu))$, which we use in the calculation of the EOS.

Note that while this describes $\phi$ in equilibrium, for $n_e < n_e^c = n_e(\mu_c,m_e)$, rigorously defined as
\begin{equation}
    \left.\dv[2]{V_\mathrm{eff}(\phi,n_e^c)}{\phi}\right|_{\phi=0} = \left.\dv[2]{V_\mathrm{eff}(\phi,\mu_c)}{\phi}\right|_{\phi=0} = 0,
\end{equation}
a potential barrier prevents the scalar field from moving away from $0$. 
Here $V_\mathrm{eff}(\phi,n_e) = 1/2 \, m_\phi^2 \phi^2 + \varepsilon_e(n_e,m_e(\phi))$ is the effective potential at constant number density. 
Its minimum at a given number density (so again the solution of the scalar equation of motion) also defines $\phi(n)$ as discussed in the main text.
In general, the critical density $n_c$ is larger than the density where $\phi_\mathrm{eq} = \phi(\mu)$ starts to deviate from $0$.
Tunnelling into the equilibrium state is extremely suppressed due to the large field excursions, so if central density of a star is below the critical one, the field will remain at $0$.
This is analogous to supercooling in an ordinary phase transition.
On the other hand, once the critical density is surpassed somewhere within the star, $\phi$ will indeed globally track its equilibrium solution.

Consequently, there are effectively two EOS to consider, a stable one where simply $\phi = \phi_\mathrm{eq}$ as well as a metastable one where $\phi=0$.
The latter is only valid as long as the maximum density in the star is below $n_e^c$ and it is simply described by the EOS without any scalar field contributions.

\subsection{Lighter version of the QCD axion} \label{app:ZnAxion}
Another well-motivated example for a light scalar field quadratically coupled to the Standard Model is the QCD axion $a$.
The defining $a G\tilde{G}$ coupling, where $G$ and $\tilde{G}$ are the gluon field strength and its dual, necessary to solve the strong CP-problem, at low energies, leads to a quadratic axion-nucleon coupling of the form \citep{Hook:2017psm,Springmann:2024mjp}
\begin{equation}
\mathcal{L}\simeq-\sigma_N\bar{N}N\left[\sqrt{1-\frac{4z}{(1+z)^2}\sin^2\left(\frac{a}{2f_a}\right)}-1\right]. \label{eq:AxionNucleonCoupling}
\end{equation}
Here $N = (n, p)$ is the nucleon-doublet, $f_a$ the axion-decay constant, $z=m_u/m_d$ is the up to down quark mass ratio, and $\sigma_N \simeq 50~\mathrm{MeV}$ the nucleon sigma term.
In the same way as described in the case of a scalar field coupling to electrons, this coupling can source the axion in dense matter as shown in \cite{Hook:2017psm,Balkin:2022qer,Balkin:2023xtr}.
For a normal QCD-axion, where the height of the potential, or equivalently, the mass at given $f_a$, is fully fixed, this would only occur above nuclear saturation density, where one lacks perturbative control.
If the height of the potential was lower or the mass of the axion lighter at a given $f_a$, sourcing occurs at lower densities and can be probed and constrained in white dwarfs \citep{Balkin:2022qer}.
In particular, this can be achieved in a technically natural way with $\Ncurl$ copies of the Standard Model and enforcing a $\mathbb{Z}_\Ncurl$- symmetry as first described by \cite{Hook:2018jle}.
Requiring the strong CP-problem to still be solved fixes $\Ncurl$ to be odd.
For large $\Ncurl$, the axion potential than can be approximated as \citep{DiLuzio:2021pxd}
\begin{equation}
   V_\Ncurl(a) \simeq \frac{m_\pi^2 f_\pi^2}{\sqrt{\pi}} \sqrt{\frac{1+z}{1-z}} \Ncurl^{-1/2} (-1)^\Ncurl z^\Ncurl \cos(\Ncurl\frac{a}{f_a}), \label{eq:ZnPotential}
\end{equation}
where $m_\pi$ and $f_\pi$ are the pion mass and decay constant, respectively.
In particular, the axion mass (at a given $f_a$) and thus the critical density are suppressed by a factor of roughly $z^\Ncurl \simeq (1/2)^\Ncurl$ with respect to the standard QCD-axion.
As shown in \cite{Balkin:2022qer}, values of $\Ncurl\geq33$ lead to a new ground state of matter in white dwarfs and come with a large gap in the predicted radii of white dwarfs, which is incompatible with observations.
For $\Ncurl = 29$ and larger, the critical density is high enough for the metastable branch to explain all observed white dwarfs, and no exclusion can be placed.
This leaves $\Ncurl=31$, which we focus on in this work.

In contrast to the simple scalar case discussed above, the potential landscape is slightly more complicated here. 
At zero density, the potential is $\frac{2\pi}{\Ncurl}$-periodic (Eq.~\eqref{eq:ZnPotential}), while the finite density contribution is only $2 \pi$ periodic (Eq.~\eqref{eq:AxionNucleonCoupling}).
Since the nucleons stay non-relativistic in a white dwarf, the finite density effective potential can simply be calculated as
\begin{equation}
    V_\mathrm{eff}(a) = V_\Ncurl(a) + n_\mathrm{Nuc} \Delta m_N(a),
\end{equation}
where $ n_\mathrm{Nuc}$ is the combined proton and neutron density and $\Delta m_N(a)$ can simply be read of from \eqref{eq:AxionNucleonCoupling}.
 
\begin{figure}
    \centering
    \begin{minipage}{0.95\linewidth}
    \centering
    \includegraphics[width=\linewidth]{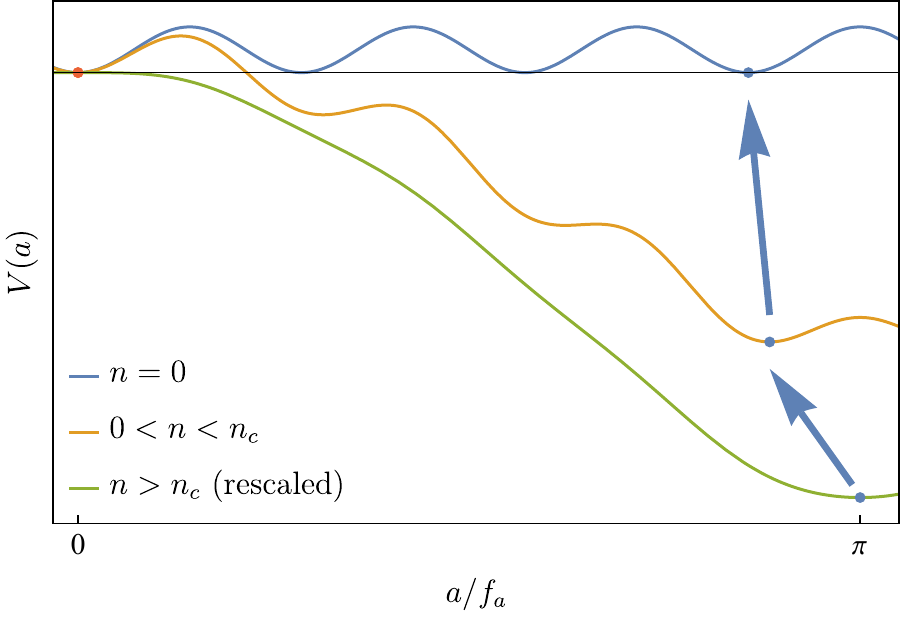}
    \end{minipage}
    \begin{minipage}{\linewidth}
    \centering
        \includegraphics[width=\linewidth]{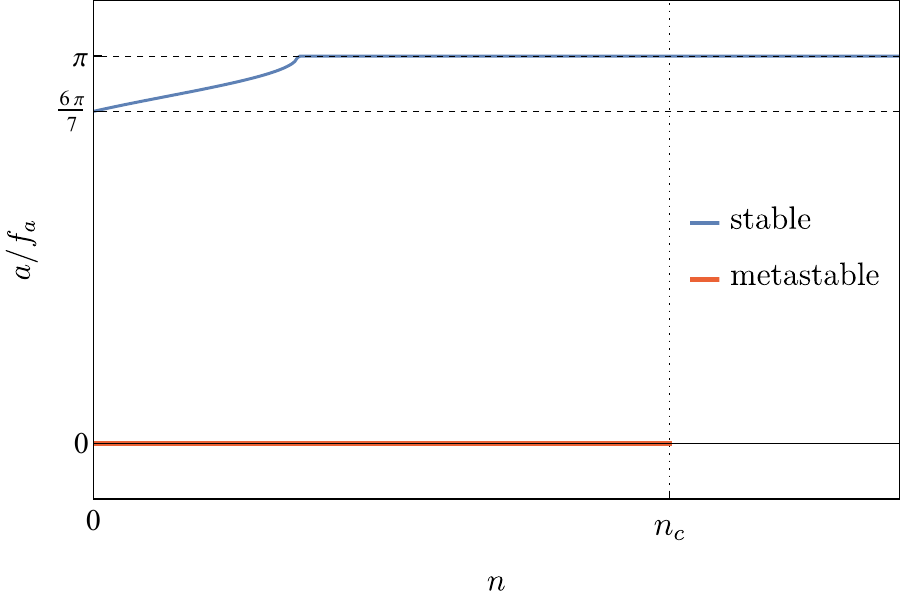}
    \end{minipage}
    \caption{Top: Potential of the $\mathbb{Z}_\Ncurl$ axion at zero and finite density.
    The arrows show the evolution of the axion expectation value along the stable branch as the density is decreased. 
    For clarity of presentation, we show $\Ncurl=7$ here, the picture stays qualitatively the same for $\Ncurl=31$.
    Bottom: Evolution of the field values on the stable and metastable branch.}
    \label{fig:ZnPotential}
\end{figure}

The potential as well as the evolution of the field value are shown in Fig.~\ref{fig:ZnPotential}.
At zero density, all minima lie at the same level (Fig.~\ref{fig:ZnPotential}, top). 
As the density is increased, the minima closer to $\pi f_a$ become lower in absolute height.
Additionally, the relative depth of the minima also becomes lower, and they start to disappear.
The last that vanishes is always the one at $a=0$, which turns into a maximum at $n=n_c$.

From the behaviour of the potential, we can now understand the behaviour of the field values and consequently the EOS.
Starting from the unsourced case, the axion field sits at $0$.
Due to the potential barrier, the axion field remains in the metastable minimum at $0$ until it fully vanishes at the critical density.
This leads to a metastable branch of the EOS and is shown in red in Fig.~\ref{fig:ZnPotential}.
At densities above the critical density, the axion field can simply track the global minimum of the finite density effective potential at $a=\pi f_a$. 
If the axion has been sourced once within a star, it will continue to track the global minimum of the potential even at densities below $n_c$.
For densities below the critical density, this global minimum at first stays at $\pi f_a$ and then continuously shifts towards the zero-density minimum closest to $\pi f_a $, at $\pi \left(1-\frac{1}{\Ncurl}\right) f_a$ as marked by the arrows in Fig.~\ref{fig:ZnPotential}.
Only at (near) zero-density, when the energy gain from the potential becomes lower than the gradient energy required for the transition to the true minimum, does the field return to the outside value of $a=0$.

From the evolution of the field values, the EOS can straightforwardly be calculated, as already described in equation~\eqref{eq:EOS_Scalar}, with the simple replacement that now the nucleon instead of the electron mass is field dependent.
In particular, for $\Ncurl\geq 31$, one finds that the branch where the scalar field sits close to $\pi$ always fulfills the macroscopic stability condition $\dv{\rho}{n}>0$, e.g., this stable branch corresponds to a cross-over type phase transition.
Meanwhile, due to the more complicated phase structure and the potential barrier around $0$, there is still a metastable branch present, overall leading to the EOS shown in Fig.~\ref{fig:EOS_ZN}.
For $\Ncurl<31$, the critical density is too high to lead to sourcing in white dwarfs and all of the observed ones are explained from the metastable branch.

%%%%%%%%%%%%%%%%%%%%%%%%%%%%%%%%%%%%%%%%%%%%%%%%%%

% Don't change these lines
\bsp	% typesetting comment
\label{lastpage}
\end{document}